\documentclass[11pt,a4paper]{article}

\usepackage[T1]{fontenc}
\usepackage[utf8]{inputenc}
\usepackage[english]{babel}
\usepackage{amsmath,amssymb}
\usepackage{graphicx}
\usepackage{geometry}
\IfFileExists{microtype.sty}{\usepackage{microtype}}{}
\usepackage{cite}
\usepackage{url}

\usepackage[
    colorlinks=true,
    linkcolor=blue,
    citecolor=blue,
    urlcolor=blue,
    bookmarks=true,
    bookmarksnumbered=true
]{hyperref}
\hypersetup{
    pdftitle={Solving the traveling salesman problem using quantum devices: QUBO and HOBO formulations},
    pdfauthor={G.V. Shuvalov, E.N. Krivtsova, M.A. Remnev, A.V. Kapranov},
    pdfsubject={Quantum optimization, traveling salesman problem},
    pdfkeywords={QUBO, HOBO, TSP, quantum computations, VQE, QAOA}
}

\title{\textbf{Solving the traveling salesman problem using quantum devices: QUBO and HOBO formulations}}

\author{G.V.~Shuvalov\thanks{e-mail: glebshu@gmail.com} \and E.N.~Krivtsova  \and A.V.~Kapranov \and M.A.~Remnev}
\date{}

\begin{document}

\maketitle

\begin{abstract}
The traveling salesman problem (TSP) forms the foundation for many logistics and optimization tasks. At the same time, this problem is NP-hard and requires significant computational resources. Quantum computing promises to substantially accelerate computations or increase solution accuracy. The TSP can be conveniently mapped onto the quadratic unconstrained binary optimization (QUBO) formulation, which in turn can be solved using variational quantum eigensolver (VQE) and quantum approximate optimization algorithm (QAOA) methods, but this mapping requires a quadratic number of qubits. The higher-order binary optimization (HOBO) formulation requires significantly fewer qubits, but it contains higher-order terms. In our work, we thoroughly investigate both formulations and provide a practical recipe for solving the TSP. We give a comprehensive study and compare in detail QUBO and HOBO formulations.
\end{abstract}


\section{Introduction}

The Travelling Salesman Problem (TSP), a quintessential NP-hard challenge in combinatorial optimization, remains a critical testbed for algorithmic innovation because of its wide-ranging applications in logistics, circuit design, and data analysis~\cite{Matai2010}. Despite its theoretical significance, exact solutions for large-scale instances---those involving hundreds or thousands of cities---are out of reach classical computers, owing to the exponential runtime of the Bellman--Held--Karp algorithm~\cite{HeldKarp1962} and other dynamical programming algorithms~\cite{Benhida2017}.

This computational bottleneck has spurred intense interest in quantum computing~\cite{Shamsuddoha2025}, especially variational algorithms like the Quantum Approximate Optimization Algorithm (QAOA)~\cite{Farhi2014} and the Variational Quantum Eigensolver (VQE)~\cite{Tilly2022}, which promise to exploit quantum parallelism to obtain approximate solutions. Many optimization and logistics problems~-- including TSP~-- can be expressed in Quadratic Unconstrained Binary Optimization (QUBO) framework~\cite{Glover2022,Lucas2014} and solved on a quantum annealer or an universal quantum computer via QAOA or VQE.

However, the current Noisy Intermediate-Scale Quantum (NISQ) era imposes stringent constraints: limited qubit counts, circuit depth restrictions, and high error rates~\cite{Preskill2018}. These limitations force a trade-off between problem fidelity and hardware scalability. Traditional formulations, such as the QUBO~\cite{Glos2022}, exacerbate this challenge because they require $N^{2}$ qubits for $N$ cities---a prohibitive demand as problem sizes grow~\cite{Lucas2014}. To address this, the Higher-Order Binary Optimization (HOBO) formulation~\cite{Glos2022} emerges as a viable alternative, reducing qubit requirements to $N \times \log(N)$ through binary encoding. Yet this efficiency comes at a cost: HOBO introduces higher-order terms in the Hamiltonian, necessitating complex multi-qubit interactions that strain variational algorithms such as QAOA (and, by extension, VQE).

Previous studies have highlighted both the advantages and limitations of HOBO formulations in the quantum-based optimization problem. HOBO naturally encodes global constraints and can improve the quality of QAOA solutions relative to QUBO, although it typically demands deeper circuits and is more sensitive to noise on real hardware~\cite{Glos2022,Ramezani2024,Schnaus2024,Bell2025}. In particular, experiments that employed Hamiltonians containing with three-qubit terms compared QAOA on IBM and quantum annealing (QA) on D-Wave. The QA on noisy devices often outperformed QAOA because the latter suffers from error accumulation~\cite{Pelofske2024,Pelofske2023,Gilbert2023}. For QAOA, it is shown that error thresholds must be respected in order to compete with QA on HOBO-encoded tasks, and that topological techniques such as factorization of two-qubit operations can dramatically reduce the number of CNOTs and increase the feasibility of the algorithm~\cite{Bell2025,Verchere2023}.

In addition to fundamental models such as factorization, hypergraph problems, logistics, and games~\cite{Jun2023,Gilbert2023,Ali2025}, several works have described algorithms and landscapes analyses for HOBO and QUBO encodings of routing and traveling salesman problems. For the latter, analytical and numerical implementations QAOA, VQE, and Grover are reported, achieving exact solutions for up to 6 cities and demonstrating that a HOBO encoding can attain the same performance with fewer circuit layers~\cite{Sato2025,Schnaus2024}. An efficient two-step algorithm for extracting a tour from the HOBO solution was also proposed. Moreover, it has been observed that qubit-based formulations can be directly mapped onto HOBO circuits~\cite{Vargas2021}. Thus, this body of work underscores the need for further development and systematic evaluation of HOBO approaches, especially for TSP, which is the focus of the present study.

In this work, we systematically investigate the solution of the TSP in the HOBO formulation and benchmark it against the QUBO formulation. Both the QAOA and the VQE algorithms are studied. We evaluate two performance metrics: the ratio of the obtained tour cost to the optimal cost, and the percentage of instances for which the exact optimal tour is recovered. In addition, we study how the number of Hamiltonian terms scales with the number of two-qubit gates, thereby quantifying the circuit depth overhead associated with each encoding.

\section{QUBO/HOBO formulations in the TSP problem}

\subsection{QUBO and HOBO formulations}

Quadratic unconstrained binary optimization (QUBO) formulates a pseudo-Boolean cost function $f:\mathbb{B}^{N}\rightarrow\mathbb{R}$ as a quadratic polynomial over $n$ binary $x_{i}\in\mathbb{B}$, where $i = 1,\ldots,N$ and $\mathbb{B} = \{0,\,1\}$:
\begin{equation}
f_{QUBO} = \sum_{i = 1}^{N}c_{i}x_{i} + \sum_{i < j}^{N}c_{ij}x_{i}x_{j}
\label{eq:fqubo}
\end{equation}
where $c_{i}$ and $c_{ij}$ are the real-valued linear and quadratic coefficients, respectively. Formally, we say that the objective of the QUBO is to calculate the binary string $x^{*}$ which minimizes the cost function $f_{QUBO}$
\begin{equation*}
x^{*} = \arg\min_{x \in \mathbb{B}^{N}}f_{QUBO}.
\end{equation*}

Higher-order binary optimization (HOBO) is an extension of QUBO and is represented by a polynomial containing terms of degree three or higher. A typical HOBO objective reads
\begin{equation}
f_{HOBO} = \sum_{i = 1}^{N}c_{i}x_{i} + \sum_{i < j}^{N}c_{ij}x_{i}x_{j} + \sum_{i < j < k}^{N}c_{ijk}x_{i}x_{j}x_{k}.
\label{eq:fhobo}
\end{equation}

Note that in general the cost function can have a constant term, which has no influence on $x^{*}$. Here and further, we do not pay any attention to the constant term in $f_{QUBO}$ and $f_{HOBO}$.

To implement the problem formulated in QUBO/HOBO approach on a quantum computer, each variable $x_{i}$ is replaced by
\begin{equation}
x_{i} = \frac{1}{2}(1 - Z_{i}),
\label{eq:xtoz}
\end{equation}
where $Z_{i}$ is the Pauli-$Z$ operator acting on the $i$-th qubit. In this way, combinatorial optimization problem is represented as a Hamiltonian to the state of qubits.

\subsection{TSP in QUBO/HOBO approach}

\subsubsection{TSP as QUBO}

As mentioned above, the TSP is a well-studied NP-hard combinatorial optimization problem with known approximation results~\cite{Christofides2022}. Formally, TSP is defined as follows: given a weighted graph with $n$ nodes (representing locations), find a cycle such that every node is visited exactly once at each timestep $t \in \{1,\ldots,\,N\}$ with the minimum sum of weights of edges used (the distances between locations). This is also known as a Hamiltonian cycle (minimum weighted cycle). In order to create the QUBO for TSP, we use a one-hot encoding to denote the index of the node visited~\cite{Lucas2014}. Therefore, the binary variable $x_{it}$ is the indicator of the condition if the $i$-th city is visited at timestep $t$. If the condition is true, then $x_{it} = 1$, and otherwise $x_{it} = 0$. Thus, we encode a valid route as a minimizing the following cost Hamiltonian
\begin{equation}
H_{cost} = \sum_{i,j}^{N} w_{ij} \sum_{t = 1}^{N - 1} x_{t,i}\, x_{t + 1,j},
\label{eq:hcost-qubo}
\end{equation}
where $w_{ij}$ is an element of the cost matrix $w$ with size $n \times n$, which means length between the $i$-th and $j$-th cities.

\begin{figure}[htbp]
    \centering
    \includegraphics[width=\textwidth]{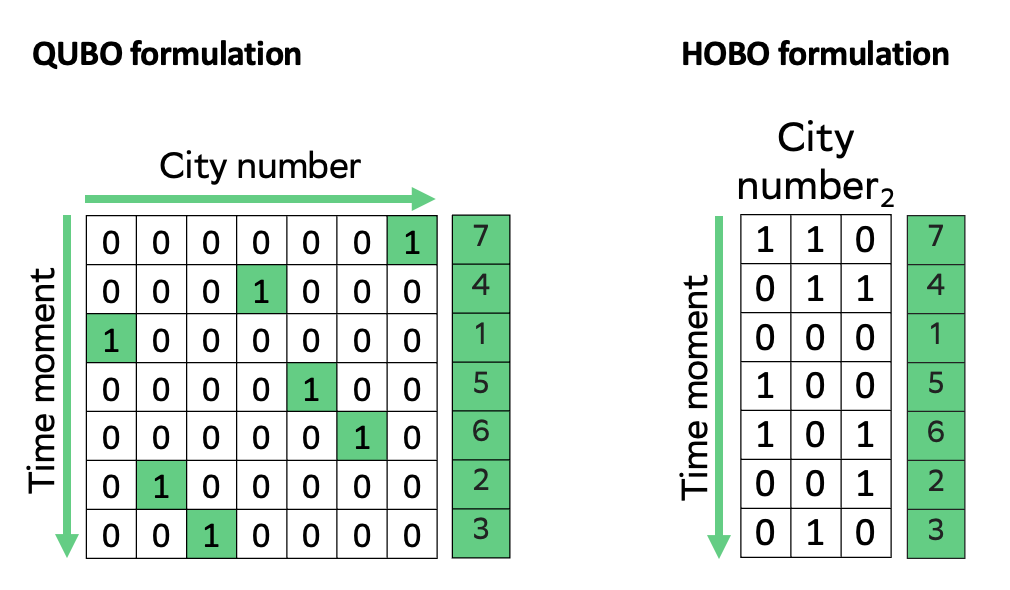}
    \caption{Example solutions of TSP in QUBO and HOBO formulations. The resulting solution matrices ultimately define the same path. In practice, data matrices may be filled in differently, but they will be solution matrices in the following cases: QUBO~-- a matrix can be a solution only if it is a permutation matrix; HOBO~-- a matrix can be a solution only if it has no repeating rows and each row specifies a city number that is no greater than the total number of cities.}
    \label{fig:solutions}
\end{figure}

TSP is natively defined over the permutations of $\{1,\ldots,N\}$ and we store all bits into $N$ collections $x_{t}$, where each collection encodes a city visited in particular timestep $t$. Thus, each city is represented by one-hot vector. The columns of the cost matrix $w$ correspond to the cite number and rows, respectively, the time step (see Fig.~\ref{fig:solutions}). Note that any route can be represented in $N$ different ways, depending on which city is visited at time $t = 1$. Therefore, we need $N^{2}$ qubits to describe the TSP.

This encoding has a well-known disadvantage: there are solutions, which do not correspond to a valid TSP route. We face with the case when two collections encode the same city. It can be caused because the several cities are visited at the same timestep or the same cities are visited multiple times in a route. These two conditions result in two separate sets of one-hot constraints, which can be written as follows
\begin{equation}
H_{constraint} = P\sum_{i = 1}^{N}\left(1 - \sum_{t = 1}^{N}x_{t,i}\right)^{2} + P\sum_{t = 1}^{N}\left(1 - \sum_{i = 1}^{N}x_{t,i}\right)^{2},
\label{eq:hconstraint-qubo}
\end{equation}
where $P$ is a penalty factor.

Thus, the complete Hamiltonian for TSP in QUBO representation takes form
\begin{equation}
H_{QUBO} = H_{cost} + H_{constraint}
\label{eq:hqubo}
\end{equation}

\subsubsection{TSP as HOBO}

Another framework is the one in which each vector $b_{t}$ encodes a city in a binary number instead of a one-hot way (see Fig.~\ref{fig:solutions}). This approach results in reducing the number of necessary qubits and is known as HOBO~\cite{Glos2022}. To store each binary number $x_{t}$ required $K = \left[\log_{2}N\right]$ qubits. It results in $N\left[\log_{2}N\right]$ qubits necessary to encode a route.

Due to such encoding we need to compare two binary numbers for equivalence. Thus, there is a new term in Hamiltonian $H_{\delta}$, which is constructed to give a zero output when the inputs are the same. This term is as follows~\cite{Glos2022}:
\begin{equation}
H_{\delta}(a,b) = \prod_{k = 1}^{K}\left(1 - \left(a_{k} - b_{k}\right)^{2}\right),
\label{eq:hdelta}
\end{equation}
where $a_{k}$ and $b_{k}$ are binary numbers. This function records the exact timesteps at which cities are visited. Note that in case of QUBO we took $H_{\delta}\left(b_{t},\,\mathrm{binary}(i)\right) = b_{ti}$, where $\mathrm{binary}(i)$ is a binary representation of $i$.

The cost Hamiltonian in binary representation takes the form~\cite{Glos2022}:
\begin{equation}
H_{cost} = \sum_{i,j}^{N} w_{ij} \sum_{t = 1}^{N - 1} H_{\delta}\left(b_{t},\,\mathrm{binary}(i)\right) H_{\delta}\left(b_{t + 1},\,\mathrm{binary}(j)\right),
\label{eq:hcost-hobo}
\end{equation}
where $w_{ij}$ is an element of the cost matrix $w$ with size $n \times K$.

Now we need to rewrite constraint conditions in such encoding. The first one we already have because we know the way of comparing two binary numbers by term $H_{\delta}$. Thus, we get the same constraint as for QUBO when two collections encode the same city (see Eq.~\eqref{eq:hconstraint-qubo})
\begin{equation}
H_{constraint} = P\sum_{t = 1}^{N - 1}\sum_{T = t + 1}^{N} H_{\delta}\left(b_{t},\,b_{T}\right).
\label{eq:hconstraint-hobo}
\end{equation}

The second raises directly from the binary encoding. We need to define the $H_{valid}$ term which ensures that the binary index for a timestep does not exceed the number of nodes (cities) in the graph. Then this term takes the form
\begin{equation}
H_{valid} = P\sum_{t = 1}^{N}\sum_{k = N + 1}^{2^{K}} H_{\delta}\left(b_{t},\,\mathrm{binary}(k)\right).
\label{eq:hvalid}
\end{equation}

A detailed proof can be found in Supplementary materials in~\cite{Glos2022}.

Thus, the total Hamiltonian for TSP in HOBO representation is the sum of all terms
\begin{equation*}
H_{HOBO} = H_{cost} + H_{constraint} + H_{valid}
\end{equation*}

\section{Methods}

\begin{figure}[htbp]
    \centering
    \includegraphics[width=\textwidth]{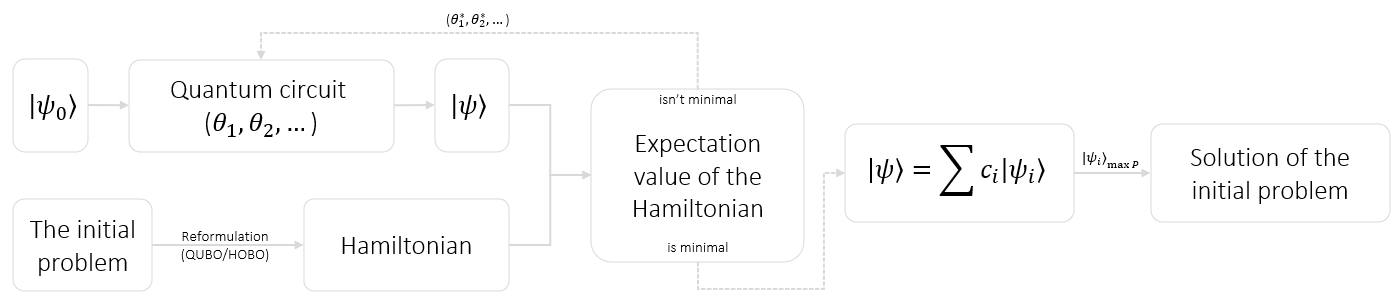}
    \caption{General structure of variational quantum algorithms. In short, a parameter-dependent quantum circuit acts on the qubit register in the initial state, producing a new quantum state from which the energy is calculated. Consequently, the expectation value of the Hamiltonian is a function of the circuit parameters.}
    \label{fig:vqa}
\end{figure}

To find the minimum energy and the corresponding qubit state of a target Hamiltonian, several approaches can be employed. In this work the minima are obtained with variational quantum algorithms (VQAs)~\cite{Qi2024}, which are executed on universal quantum devices (in our case on noise-free simulators). The structure of a VQA is shown in Fig.~\ref{fig:vqa}. A VQA is a hybrid quantum-classical algorithm. The quantum circuit is parametrized by a set of angles $\theta_{1},\ \theta_{2},\ \theta_{3},\ldots$. After circuit is applied, the resulting wave function depends on these parameters $\psi(\theta_{1},\ \theta_{2},\ \theta_{3},\ldots)$. Then, qubit measurements are performed, or the expectation value of the Hamiltonian is evaluated. The classical optimizer treats the measured expectation value as a cost function and updates the parameters $\theta_{i}$ so as minimize it, for example be gradient descent.

The quantum circuits employed for the VQA are the QAOA and Hardware-Efficient ansatz for VQE (VQE-HEA). Below we discuss each of them in more detail.

QAOA is an algorithm that simulates an adiabatic change of an initial Hamiltonian into a target Hamiltonian~\cite{Farhi2014}. In this situation, theory predicts that the lowest-energy state of the initial Hamiltonian should evolve into the lowest-energy state of the target Hamiltonian (in our case the QUBO or HOBO Hamiltonian). For QAOA the conventional choice of initial Hamiltonian is $\widehat{H} = -\sum_{i = 1}^{N_{q}}\widehat{X}_{i}$, where $N_{q}$~-- is the total number of qubits employed in the calculation. Ground-state of this Hamiltonian is $|++\ldots+\rangle$.

To model this process, the Suzuki--Trotter decomposition is employed (in its simplest, first-order form), which approximates the evolution operator of a Hamiltonian with non-commuting terms by sequentially applying the evolution operators of the individual parts of that Hamiltonian:
\begin{equation}
e^{-i(A + B)t} = \lim_{N \rightarrow \infty}\left(e^{-iAt/N}e^{-iBt/N}\right)^{N}
\label{eq:trotter}
\end{equation}

QAOA essentially performs a similar operation, but instead of using equal time intervals and annealing schedules it employs implicit parameters no longer have an explicit physical meaning, they are simply tuned by some minimization method:
\begin{equation}
|\psi\rangle = e^{-i\gamma_{d}H_{init}}e^{-i\beta_{d}H_{fin}}\ldots e^{-i\gamma_{1}H_{init}}e^{-i\beta_{1}H_{fin}}|\psi_{0}\rangle
\label{eq:qaoa}
\end{equation}

Note that in the limit of infinitely many layers, QAOA essentially performs the same operation as the Suzuki--Trotter decomposition.

\begin{figure}[htbp]
    \centering
    \includegraphics[width=\textwidth]{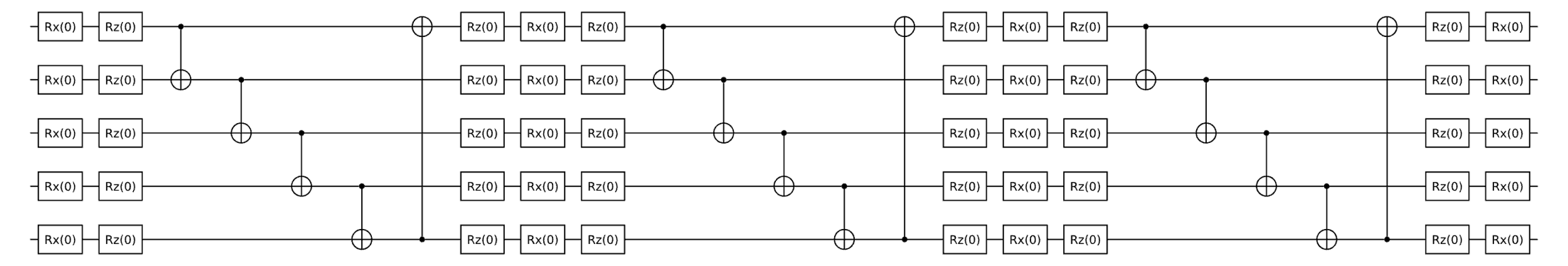}
    \caption{HEA-ansatz for VQE, using in this work (all parameters equal 0, all parameters are independent).}
    \label{fig:hea}
\end{figure}

VQE with HEA is a classic example of a variational quantum algorithm (see Fig.~\ref{fig:hea}). This ansatz is inspired more by classical algorithms than by any particular quantum system. Each of its layers consists of a set of single-qubit gates followed by an entangling layer that connects all qubits in a chain-like topology. Overall, the construction is motivated by the architecture of classical neural networks.

The VQE-HEA architecture does not depend on particular problem Hamiltonian, and can therefore be employed unchanged for both QUBO-type and HOBO-type problems.

\begin{figure}[htbp]
    \centering
    \includegraphics[width=\textwidth]{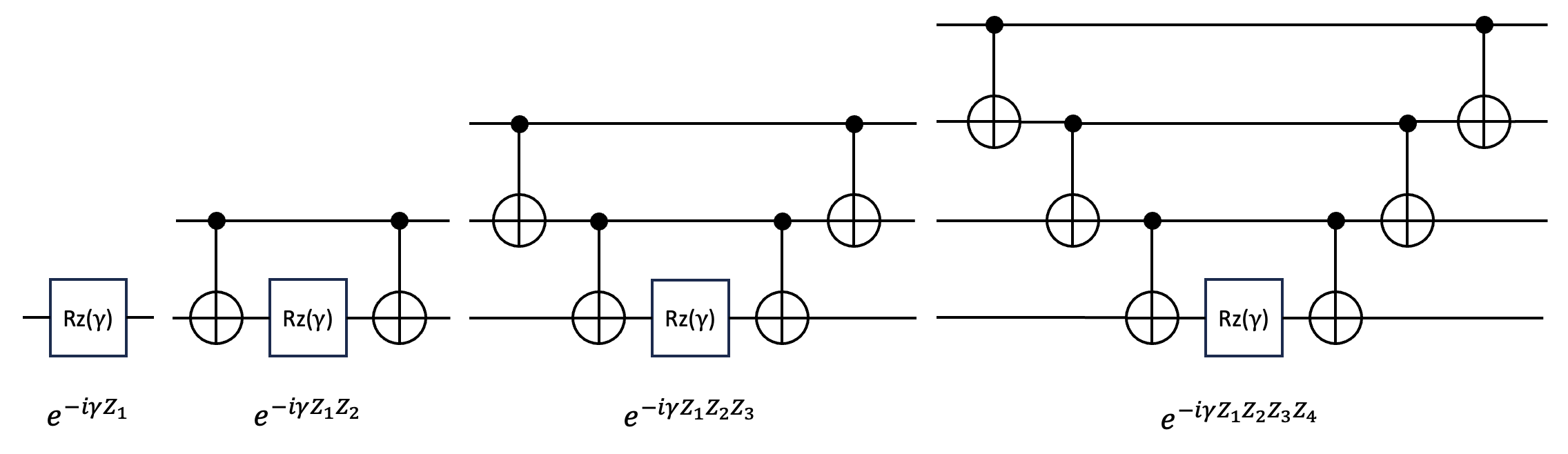}
    \caption{Example of decomposing Trotter-expanded terms into quantum Clifford gates. Further decompositions are constructed by analogy.}
    \label{fig:decomposition}
\end{figure}

QAOA, in contrast, depends on the form of the problem Hamiltonian, so for QUBO and HOBO formulations the quantum circuit will differ somewhat. The difference is that in a QUBO formulation only single-qubit and pairwise interactions appear, whereas a HOBO formulation contains higher-order interactions. Consequently, the decomposition operators of the form $e^{-i\alpha Z_{i}Z_{j}Z_{k}\ldots}$ must be generalized. One possible decomposition of such a multi-qubit operator is shown in Fig.~\ref{fig:decomposition}.

The optimizer employed is Adaptive Moment Estimation (ADAM)~\cite{Kingma2017}. Gradients are computed using a second-order finite-difference formula for QAOA and the parameter-shift rule for VQE (since the gate parameters are independent)~\cite{Wierichs2022}.

The quantum-simulation frameworks used were Yao.jl~\cite{Luo2020} and NVIDIA CUDA-Q~\cite{CUDAQ2025}. The first framework was employed for many small-scale runs because of its efficient gradient computation; whereas the latter was used for larger simulations owing its effective utilization of GPU resources.

The main metrics used to compare the QUBO and HOBO formulations were:
\begin{itemize}
\item Approximation ratio (AR)~-- the ratio of the energy obtained by the algorithm to the globally minimal energy. In the figures we plot the inverse of this quantity (so that the value lies between 0 and 1); larger values indicate better performance.
\item True percentage (TP)~-- the percentage of correct answers among 10000 measurements of the quantum circuit. Larger values are better.
\end{itemize}

\section{Results}

The generators for QUBO and HOBO Hamiltonians were prepared as symbolic expressions, using SymPy~\cite{Meurer2016} for Python and Symbolics.jl~\cite{Gowda2022} for the Julia, to facilitate subsequent work with them. Because the Hamiltonians were originally formulated in binary form, the transformation~\eqref{eq:xtoz} had to be applied to convert the binary variables into unitary ones.

\begin{figure}[htbp]
    \centering
    \includegraphics[width=\textwidth]{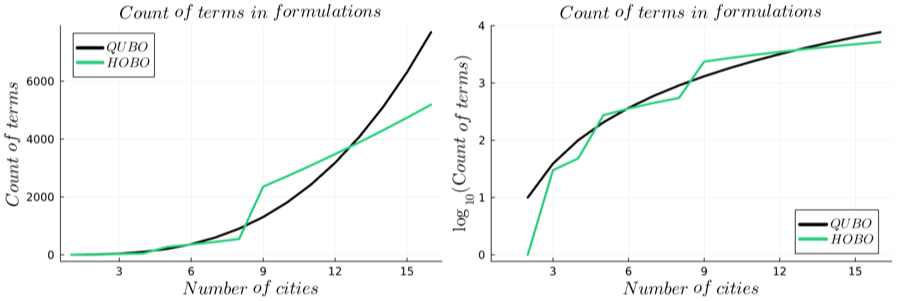}
    \caption{Count of terms in formulations QUBO and HOBO (the number of terms is calculated for unitary forms of these Hamiltonians). Right image is depicted in logarithm scale.}
    \label{fig:terms}
\end{figure}

Fig.~\ref{fig:terms} shows plots of the number of terms in the unitary QUBO and HOBO Hamiltonians, which is important, for example, when using them in QAOA. It can be seen~-- especially clearly on the logarithmic plot~-- that the number of terms grows asymptotically the same for both formulations. An analytical formula for the number of terms in the QUBO formulation is known:
\begin{equation}
f(N) = 2N^{3} - 2N^{2} + N,
\label{eq:terms}
\end{equation}
where $N$ is the number of cities in the problem instance.

\begin{figure}[htbp]
    \centering
    \includegraphics[width=\textwidth]{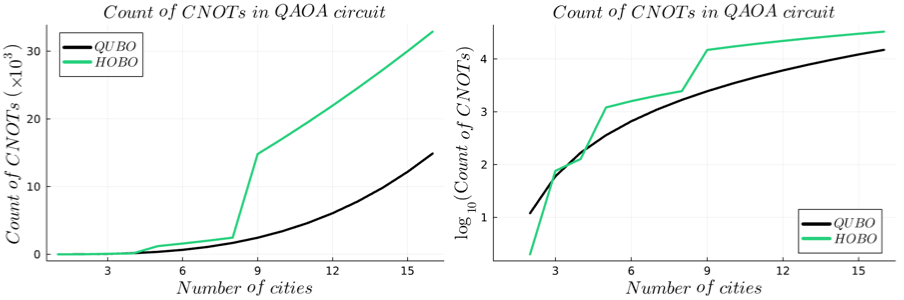}
    \caption{Count of CNOTs in QAOA layer, built using QUBO and HOBO Hamiltonians. Right image is depicted in logarithm scale.}
    \label{fig:cnots}
\end{figure}

Another important characteristic is the number of CNOT gates in the QAOA circuit, because the decomposition of the trotterized terms differ in QUBO and HOBO formulations with respect to the gate count. The results are shown in Fig.~\ref{fig:cnots}. An analytical expression the number of CNOTs in the QAOA circuit for the QUBO formulation is also known:
\begin{equation}
f(N) = 4N^{3} - 6N^{2} + 2N.
\label{eq:cnots}
\end{equation}

\begin{figure}[htbp]
    \centering
    \includegraphics[width=\textwidth]{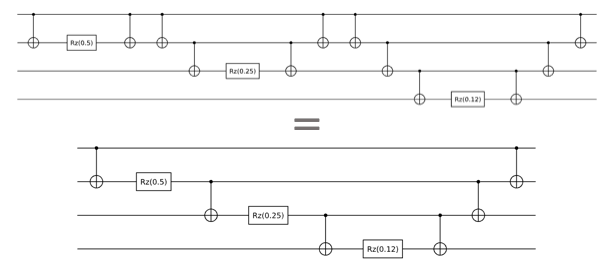}
    \caption{Circuit compression method: union of several decomposition of evolution operator (like in QAOA) into one operator, which require a smaller number of CNOTs.}
    \label{fig:compression}
\end{figure}

Overall, it can be seen that the number of CNOT gates for the HOBO formulation grows much more rapidly. This growth can be mitigated, for example, by merging several trotterized terms, as shown in Fig.~\ref{fig:compression}.

\begin{figure}[htbp]
    \centering
    \includegraphics[width=\textwidth]{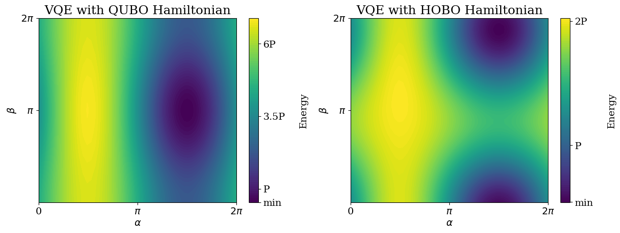}
    \caption{Landscape of the mean value of the Hamiltonian obtained by VQE, near the global minimum ($\alpha$ and $\beta$~-- randomly selected parameters; min~-- minimal energy; $P$~-- penalty coefficient). Count of cities~-- 3.}
    \label{fig:landscape-vqe}
\end{figure}

\begin{figure}[htbp]
    \centering
    \includegraphics[width=\textwidth]{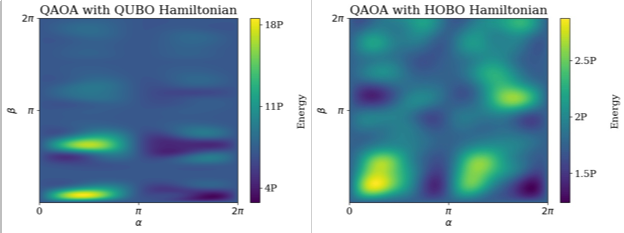}
    \caption{Landscape of the mean value of the Hamiltonian obtained by QAOA (depth of QAOA circuit equal 1, therefore $\alpha$ and $\beta$~-- all parameters of this circuit; $P$~-- penalty coefficient). Count of cities~-- 3.}
    \label{fig:landscape-qaoa}
\end{figure}

Using these Hamiltonians, one can construct the energy landscapes obtained when solving them with VQE or QAOA (Fig.~\ref{fig:landscape-vqe} and Fig.~\ref{fig:landscape-qaoa}~-- the underlying optimization problem is identical for all combinations of quantum circuits and Hamiltonians). The landscape of the HOBO formulation is more intricate than that of the QUBO formulation, although the overall energy range for the HOBO case in smaller. This occurs because penalties in the HOBO formulation are harder to obtain, owing to the overall smaller number of invalid solutions in that formulation. The reduced amplitude of energy-landscape's peaks and valleys can potentially improve optimizer performance, helping the algorithm converge more reliably toward the global minimum.

\begin{figure}[htbp]
    \centering
    \includegraphics[width=\textwidth]{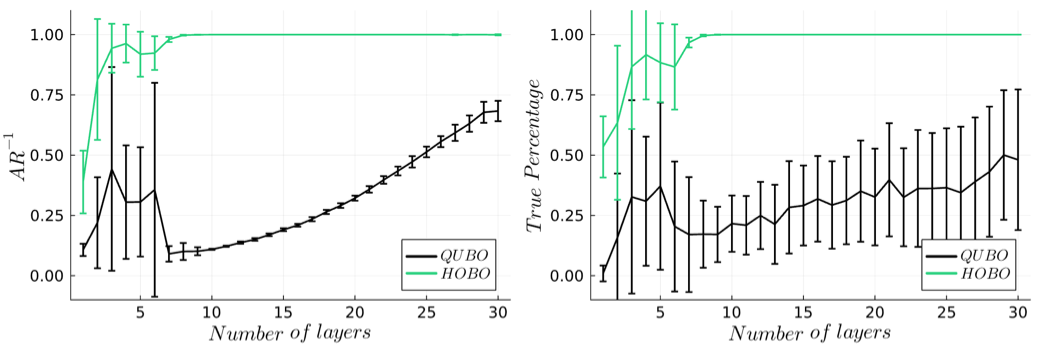}
    \caption{Dependence of $AR^{-1}$ and $TP$ on the number of VQE layers. Count of cities~-- 3, number of training iterations~-- 1000. For each depth setting, statistics were gathered from 50 problem instances, and the error was estimated from these data (the initial parameters were selected randomly in each launch).}
    \label{fig:metrics-vqe}
\end{figure}

\begin{figure}[htbp]
    \centering
    \includegraphics[width=\textwidth]{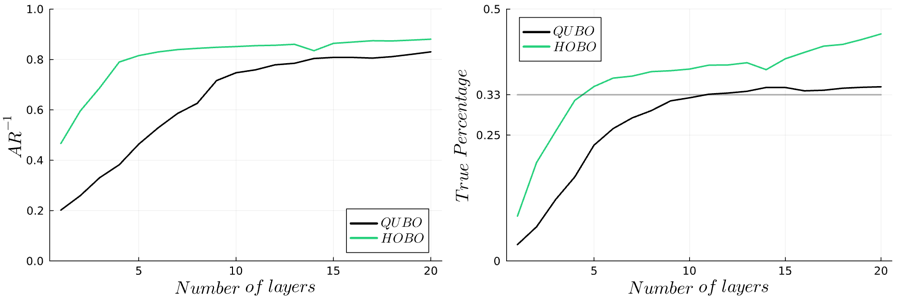}
    \caption{Dependence of $AR^{-1}$ and $TP$ on the number of QAOA layers. Count of cities~-- 3, number of training iterations~-- 1000. In the plot of TP metric, a reference line at 33\% is shown. This line is introduced because QAOA initially prepares uniform state. During the algorithm the invalid states are strongly suppressed, while solutions against the background of valid states are only weakly amplified. Consequently, the worst-case scenario the output distribution after QAOA collapses to a uniform mixture of all valid states, and the probability of obtaining a correct solution is $P_{N = 3} = \frac{2}{N!} = \frac{2}{6} = \frac{1}{3}$, where $N$~-- count of cities in TSP.}
    \label{fig:metrics-qaoa}
\end{figure}

The statement above is generally corroborated by Fig.~\ref{fig:metrics-vqe} and Fig.~\ref{fig:metrics-qaoa}, which depict how the $AR^{-1}$ and $TP$ metrics depend the number layers in the variational circuit (QAOA and VQE).

For VQE, errors were also computed because each run used random parameters, whereas in QAOA the initial parameters were set to zero.

On these figures it is clearly visible that the HOBO formulation outperforms the QUBO formulation on both metrics in both cases, ultimately converging more reliably to the correct solution.

In the case of VQE, once the circuit reached a certain number of layers, it began to reliably converge to the global minimum for both the QUBO and HOBO formulations.

\begin{figure}[htbp]
    \centering
    \includegraphics[width=\textwidth]{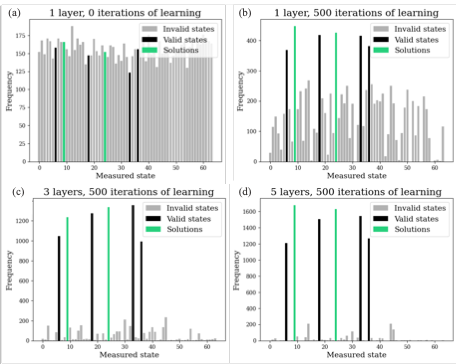}
    \caption{Circuit measurements of QAOA with some numbers of layers and training iterations, count of cities in TSP~-- 3, encoding of TSP~-- HOBO. State number is constructed as a decimal representation of the bitstring, obtained from the measured states of the system of qubit. Gray lines mark invalid solutions, black and green lines mark valid solutions, green lines mark solutions of the TSP problem. (a)~-- 1 layer of QAOA without parameter optimization, so the parameters are zero. The result state is uniform state. (b)~-- 1 layer of QAOA with parameter optimization (500 training iterations). Invalid states are suppressed. (c)~-- 3 layers of QAOA with parameter optimization (500 training iterations). Invalid states are suppressed more strongly than in case (b). (d)~-- 5 layers of QAOA with parameter optimization (500 training iterations). Invalid states are almost completely suppressed, the growth of decisions over valid states begins.}
    \label{fig:measurements}
\end{figure}

In the case of QAOA, Fig.~\ref{fig:metrics-qaoa} also includes a reference line at 33\%. This line corresponds to the fraction of solutions one would obtain if the final quantum state were a uniform mixture over all valid (all city permutations) states. The purpose of the line is to illustrate that, even with a shallow circuit depth, QAOA suppresses the invalid states quite efficiently~-- a behavior that can be seen more clearly in Fig.~\ref{fig:measurements}. As the circuit depth is increased, the proportion of valid solutions rises relative to the total number states moving the performance above the 33\% baseline.

Returning to Fig.~\ref{fig:metrics-qaoa}, one can see that the HOBO formulation surpasses the 33\% threshold rather quickly, whereas the QUBO formulation reaches the same level much more slowly. This suggests that QAOA combined with the HOBO encoding is more effective at isolating solutions among the set of valid states, but overall, QAOA is better at distinguishing valid solutions from invalid ones.

\section{Discussion}

Based on the obtained results, can conclude that HOBO formulation performs more efficiently than the QUBO formulation when solving TSP. This advantage is likely due not only to the reduction in the number of qubits required but primarily to the decrease in the number of non-existent solutions.

An additional, important advantage of the HOBO formulation~-- unlike QUBO~-- is that it constrains the eigen-energies of the states more tightly. This tighter bound arises both the smaller number of invalid solutions and from the way the Hamiltonian itself is constructed. As a result, VQE and QAOA that employ the HOBO formulation reach energies close to the exact solutions more quickly in terms of circuit depth (and, secondary, in terms of the number of training iterations).

From the perspective of the optimization landscape with respect to the variational algorithm parameters, the HOBO Hamiltonian is more intricate. However, because its landscape is much flatter, it is potentially better suited for gradient-based methods: the flatter terrain makes it easier to escape from local minimum, for example by leveraging noise, momentum, or other perturbations.

The main drawback remains that the HOBO formulation is rather bulky. Although the number of terms is comparable to that of the QUBO formulation, the number of required entangling gates (in this case CNOTs) grows faster than in the QUBO formulation. Since, in practice, the fidelity of entangling gates is lower than that of single-qubit gates, HOBO consequently loses to QUBO in this respect. However, there exist techniques~-- such as the one described in reference~\cite{Verchere2023}, that can reduce the count of two-qubit gates by order of magnitude, thereby improving the overall accuracy of the HOBO approach.

Thus, future research on the HOBO formulation of TSP should primarily focus on adapting it to real quantum hardware. In addition, exploiting QAOA's ability to efficiently filter out invalid solutions appears to be a promising direction.

\section{Conclusion}

The traveling salesman problem is a rather challenging problem to reformulate for quantum devices.

In this paper two formulations for universal quantum computers are examined the QUBO formulation and the HOBO formulation.

The first one is intrinsically simpler, but it requires a quadratic number of qubits with respect to the number of cities, whereas the second one needs only $N\left\lceil\log_{2}N\right\rceil$ qubits ($N$~-- the number of cities), which makes it potentially more efficiently stimulable on classical hardware. The main drawback of the second formulation is that, when used naively becomes larger (especially if the quantum algorithm, in some way, requires the Hamiltonian to be embedded inside the circuit). HOBO yields better results than QUBO: its convergences more efficiently toward the exact energy minimum and, as consequence, more frequently returns the correct tours. This advantage is likely related to the flatter landscape of the HOBO formulation with respect to the variational parameters of quantum algorithms. In addition, it contains fewer invalid states and requires fewer qubits.

\bibliographystyle{unsrt}
\bibliography{references}

\end{document}